%% file: main.tex
\documentclass[conference, a4paper]{IEEEtran}

\renewcommand\IEEEkeywordsname{Keywords}

\IEEEoverridecommandlockouts

\usepackage[utf8]{inputenc} %

\usepackage[T1]{fontenc} %
\usepackage{textcomp}

\usepackage[
            backref=true,
            backend=biber,
			sorting = none,
			url = false,
			doi = false,
			isbn = false,
			bibstyle=ieee, %
			citestyle=numeric-comp, %
			]{biblatex}
				
\renewcommand*{\bibfont}{\footnotesize} %
\usepackage[table,xcdraw,dvipsnames]{xcolor}
\definecolor{ZSWDarkBlue}{RGB}{0,89,170}
\usepackage[unicode=true,
            bookmarks=true,
            bookmarksnumbered=false,
            bookmarksopen=false,
            breaklinks=false,
            hidelinks,     %
            colorlinks=False]
            {hyperref}
\hypersetup{pdftitle={Influence of Demand and Generation Uncertainty on the Operational Efficiency of Smart Grids},
            pdfauthor={B. Matthiss, A. Momenifarahani, K. Ohnmeiss, M. Felder},
            citecolor = ZSWDarkBlue,
            linkcolor = ZSWDarkBlue}

\newbibmacro{string+doiurlisbn}[1]{%
  \iffieldundef{doi}{%
    \iffieldundef{url}{%
      \iffieldundef{isbn}{%
        \iffieldundef{issn}{%
          #1%
        }{%
          \href{http://books.google.com/books?vid=ISSN\thefield{issn}}{#1}%
        }%
      }{%
        \href{http://books.google.com/books?vid=ISBN\thefield{isbn}}{#1}%
      }%
    }{%
      \href{\thefield{url}}{#1}%
    }%
  }{%
    \href{http://dx.doi.org/\thefield{doi}}{#1}%
  }%
}
\DeclareFieldFormat{title}{\usebibmacro{string+doiurlisbn}{\mkbibemph{#1}}}
\DeclareFieldFormat[article,incollection]{title}%
    {\usebibmacro{string+doiurlisbn}{\mkbibquote{#1}}}

\usepackage{balance}	
\bibliography{literature}        %

\usepackage{siunitx}
\usepackage{eurosym}
\DeclareSIUnit \eur {\text{\euro}}

\newcommand*{\eurkwh}[1]{\SI[scientific-notation=fixed,fixed-exponent=0]{#1}{\text{\euro}\per{\kilo\watt\hour}}}

\usepackage{nomencl}
\makenomenclature

\usepackage[short]{optidef}

\ifCLASSINFOpdf
  \usepackage[pdftex]{graphicx}
  \graphicspath{{./figures/}}
  \DeclareGraphicsExtensions{.pdf,.jpeg,.png}
\else
\fi
\usepackage{balance}

\usepackage{ifthen}
\makeatletter
\newcounter{IEEE@bibentries}
\renewcommand\IEEEtriggeratref[1]{%
  \renewbibmacro{finentry}{%
    \stepcounter{IEEE@bibentries}%
    \ifthenelse{\equal{\value{IEEE@bibentries}}{#1}}
    {\finentry\@IEEEtriggercmd}
    {\finentry}%
  }%
}
\makeatother

\usepackage{eurosym}

\usepackage{tabularx}
\usepackage{multirow}
\usepackage{booktabs}

\usepackage[dvipsnames]{xcolor}
\usepackage{tikz} 
\usetikzlibrary{calc, angles, quotes}
\usepackage[europeanresistors,americaninductors]{circuitikz}
\usepackage{lscape}

\usepackage{mathtools} %
\usepackage{amsmath}
\usepackage{bm}
\usepackage{amssymb}
\usepackage{nomencl}
\usepackage{kbordermatrix}

\usepackage{algpseudocode}

\usepackage{optidef}

\ifCLASSOPTIONcompsoc
  \usepackage[caption=false,font=normalsize,labelfont=sf,textfont=sf]{subfig}
\else
  \usepackage[caption=false,font=footnotesize]{subfig}
\fi

\usepackage{stfloats}
\usepackage{url}
\usepackage{textcomp}
\usepackage{lipsum}

\newcommand\copyrighttext{%
  \footnotesize \textcopyright 2018 IEEE. Personal use of this material is permitted.
  Permission from IEEE must be obtained for all other uses, in any current or future 
  media, including reprinting/republishing this material for advertising or promotional 
  purposes, creating new collective works, for resale or redistribution to servers or 
  lists, or reuse of any copyrighted component of this work in other works. 
  DOI: \href{https://doi.org/10.1109/ICRERA.2018.8566733}{10.1109/ICRERA.2018.8566733}}
\newcommand\copyrightnotice{%
\begin{tikzpicture}[remember picture,overlay]
\node[anchor=south,yshift=10pt] at (current page.south) {\fbox{\parbox{\dimexpr\textwidth-\fboxsep-\fboxrule\relax}{\copyrighttext}}};
\end{tikzpicture}%
}

\begin{document}

\input{paper}

\renewcommand*{\UrlFont}{\rmfamily}
\balance
\printbibliography

\end{document}

%% file: paper.tex
\title{Influence of Demand and Generation Uncertainty on the 
Operational Efficiency of Smart Grids}

\author{\IEEEauthorblockN{B. Matthiss\IEEEauthorrefmark{1},
A. Momenifarahani\IEEEauthorrefmark{2}, K. Ohnmeiss\IEEEauthorrefmark{3} and
M. Felder\IEEEauthorrefmark{4}}
\IEEEauthorblockA{Zentrum für Sonnenenergie- und Wasserstoff-Forschung \\ Baden-Württemberg, Stuttgart, Germany\\
Email: \IEEEauthorrefmark{1}benjamin.matthiss@zsw-bw.de,
\IEEEauthorrefmark{2}arghavan.momenifarahani@stud.zsw-bw.de,\\
\IEEEauthorrefmark{3}kay.ohnmeiss@zsw-bw.de,
\IEEEauthorrefmark{4}martin.felder@zsw-bw.de}
\thanks{This work has been funded by the Federal Ministry of Economy and Energy in Germany.%
The authors gratefully acknowledge this support. They are solely responsible for the contents of this publication.}
}

\maketitle
\copyrightnotice %

\IEEEpubidadjcol

\renewcommand\nomgroup[1]{%
  \item[\bfseries
  \ifstrequal{#1}{A}{Indexes and Sets}{%
  \ifstrequal{#1}{B}{Parameters}{%
  \ifstrequal{#1}{C}{Variables}{}}}%
]}

\begin{abstract}
To ensure the smooth and near optimal operation of storage and controllable generation in a grid with a high share of renewable energies, it is important to have accurate forecasts for load and generation. But even with the advanced forecasts available today, the prediction error can have a significant impact on the operation performance of the system. This paper compares and analyzes the impact of the prediction-error on the operational performance in case of a small virtual power plant. \end{abstract}


\setlength{\nomitemsep}{0.2cm}
\nomenclature[A, 01]{$g${$\in{G}$}}{Generating units,running from 1 to $G$.} %
\nomenclature[A, 02]{$t${$\in{T}$}}{Hourly periods,running from 1 to $T$.}
\nomenclature[B, 01]{$D_{t}$}{Load demand in hour $t$.}
\nomenclature[B, 02]{$\kappa_g$}{Cost for the energy delivered by unit $g$ per time step (\$).}
\nomenclature[B, 03]{$\overline{P}_g$}{Maximum power output of unit $g$.}
\nomenclature[B, 04]{$\underline{P}_g$}{Minimum power output of unit $g$.}
\nomenclature[B, 05]{$C$}{Storage capacity.}
\nomenclature[B, 06]{$T^{D}_{min,g}$}{Minimum down-time of unit $g$.}
\nomenclature[B, 07]{$T^{U}_{min,g}$}{Minimum up-time of unit $g$.}
\nomenclature[B, 08]{$R^{U}_{g}$}{Ramp-up rate of unit $g$.}
\nomenclature[B, 09]{$R^{D}_{g}$}{Ramp-down rate of unit $g$.}
\nomenclature[C,01]{$p_{g,t}$}{Power output of unit $g$ at timestep $t$.}
\nomenclature[C,02]{$v_{g,t}$}{Binary variable which takes the value of 1 if the unit starts up and 0 otherwise.}
\nomenclature[C,04]{$u_{g,t}$}{Binary variable which is equal to 1 if the unit producing above minimum power and 0 otherwise.}
\nomenclature[C,05]{$w_{g,t}$}{Binary variable which takes the value of 1 if the unit shuts down and 0 otherwise.}
\nomenclature[C,06]{$x_t$}{Battery state of charge (SOC) at timestep $t$. $[0,1]$ }

\section{Introduction}
Governments around the world have made climate protection a priority policy goal. This is increasingly occurring as a result of climate change, and the environmental issues it has presented to us. 
To combat environmental degradation, a shift away from fossil fuels and towards renewable systems is required.
In Germany for instance, renewable forms of energy are on the rise. In 1990, renewable energy sources accounted for just under 4\% of total electrical energy production, but that share rose to as much as 33\% by 2017 \cite{kerstine_appunn_germanys_2018}. 
The number of consumers and producers who utilize renewable power generation systems has increased, which also entails new and increased challenges for system operators \cite{moreci_energy_2015,}.
When the next generation shifts increasingly towards renewable resources, extensive planning is required to meet, as well as to coordinate and control the loads of individual consumers. This well informed policy will be crucial, as considerable uncertainties can interfere when planning for energy problems.
Uncertainty from inaccurate projections of the future needs and capacities of renewable resources can also generate problems in micro grid operation. \cite{wang_hybrid_2014, athari_modeling_2016,ogimi_optimal_2012,}.

In this paper, the effects of the prediction error on the operational efficiency of a small self consumption community is investigated.

\section{Optimal Operation of Power Systems}
The motivation to apply efficient resource scheduling in power systems has increased. It supports the economical and reliable energy production \cite{scalfati_optimal_2017,anatone_comprehensive_2015,}.
It also determines the operation of generating resources.
This minimizes the total generating cost while responding to the system boundaries as well as to the demands \cite{xiaohong_guan_scheduling_1999}. 
Dynamic programming \cite{bomberger_dynamic_1966}, standard mixed integer programming as well as non-linear programming have been developed as solution methods to cope with scheduling problems in research areas.
Further techniques encompass exact approaches such as Lagrange relaxation \cite{wang_short-term_1995, xiang_yu_unit_2016}, 
or heuristics and meta-heuristics methods \cite{ting_novel_2006,dahal_modern_2000}. 
One of the most used approaches is the mixed-integer-linear programming (MILP) approach. 
Efficient solution techniques have been developed to guarantee convergence within few iterations \cite{xiaohong_guan_scheduling_1999, scalfati_optimal_2017}.
The production of large power systems is coordinated by these UC approaches.
Therefore much effort has been made to optimize the problem formulation.
This leads to a decreased number of binary variables which speeds up the process of search and limits the search space. 

\begin{figure*}[tb]
\captionsetup[subfloat]{farskip=4pt,captionskip=2pt}
\centering
\includegraphics[width=.8\textwidth]{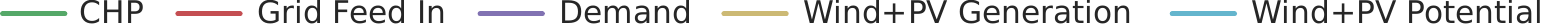}
\subfloat[\label{fig:pred_err0}]{\includegraphics[width=.4\textwidth]{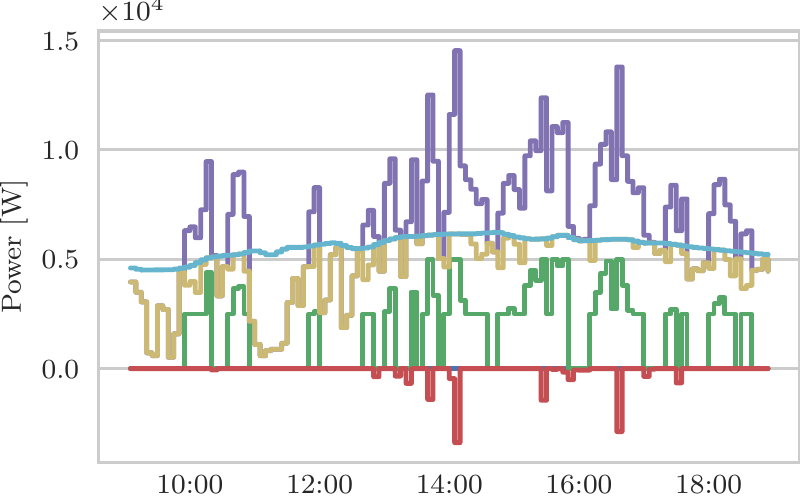}}\hfil
\subfloat[\label{fig:pred_err10}]{\includegraphics[width=.4\textwidth]{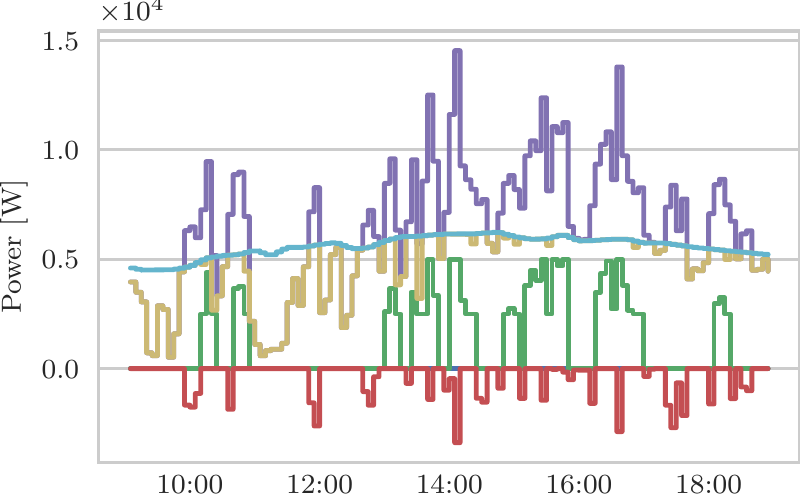}}
\caption{Exemplary results for the optimized operation. In Fig.\,\ref{fig:pred_err0} the relative prediction error is \SI{0}{\percent} and in Fig.\,\ref{fig:pred_err10} the relative prediction error is \SI{10}{\percent}. The goal of the optimization in this case is to use as little energy from the grid as possible (negative feed in, red line). }
\label{fig:results_no_batt}
\end{figure*}

In this paper the underlying optimization problem is solved in a Model Predictive Control (MPC) fashion. Hereby a dynamic model of the energy system is used to predict the future behavior within a discrete time horizon $T$. This model depends on various inputs such as the predicted generation or demand. 
At each sample point the complete optimization problem is solved over the horizon but only the first step is implemented. The software has been implemented in python using CVXPY \cite{cvxpy, cvxpy_rewriting}.
An example of the optimized operation for a small self consumption community with renewable generation and a combined heat and power (CHP) cogeneration plant is shown in Fig.\,\ref{fig:results_no_batt}.

\section{Problem Formulation} 
To standardize the modeling of renewable energy resources (PV, wind), intermittent generation units or energy storage devices with limited capacity, a standardized formulation according to the \emph{Power Node} framework is used in this paper.
The reader is referred to \cite{heussen_energy_2010} for details.
Depending on the device properties and the energy supply and demand the constraints are listed and the decision variables are defined to present the characteristic properties of power nodes.
In the case of renewable energy generation plants, it is assumed that production can only be curtailed and that otherwise there is no possibility of influence. Consumers are assumed to be passive loads, i.e. not controllable.
The cost function of the problem can be expressed as follows:

\begin{equation}
\begin{aligned}
& \underset{P_{g,t}}{\text{min}}&  \sum_{t\in T, g\in G}& P_{g,t} \kappa_g \\
& \text{s.t.}              & 0 &= \sum_{g\in G}P_{g,t}, \quad t \in T \\
&& 						\dot{x}_t &=C^{-1} p_\text{batt} \\
&&						0 &\leq x \leq 1 \\
&&				\underline{P}_\text{batt} &\leq P_\text{batt}\leq \overline{P}_\text{batt} \\
&&				0 &\leq P_\text{RES} \leq \overline{P}_\text{RES} \\
&&				\underline{P}_\text{CHP} &\leq  P_\text{CHP}   \leq \overline{P}_\text{CHP} \quad \text{if}\; P_\text{CHP}>0 \\
&& 				\underline{P}_\text{grid} &\leq  P_\text{grid} \leq \overline{P}_\text{grid} \\
\end{aligned}
\end{equation}

hereby $P_{g}$ denotes power supply/demand by generator/consumer $g$ in the network, while the $\kappa_g$ regards to the price of per kWh  supplied/consumed electricity [\EUR/kWh]. 
The variable $P_\text{grid}$ corresponds to the power delivered from the external grid and $P_\text{batt}$ describes the power generated by battery limited by the minimum power $\underline{P}_\text{batt}$ and the maximum power $\overline{P}_\text{batt}$. 
The demand of the consumers is denoted with $P_\text{dem}$ and the corresponding limit with $\overline{P}_\text{dem}$. 
Moreover, the power supply $P_\text{RES}$ is associated with the renewable energy systems (RES) with a maximum output of $\overline{P}_\text{RES}$.  The combined heat and power unit is represented by $P_\text{CHP}$ with a maximum output power $\overline{P}_\text{CHP}$ in the related network. 
The state of charge (SOC) of the battery is designated by $\dot{x}_t$ and normalized to $0 \leq x \leq 1$ with storage capacity $C$. The optimization is run over the discrete time horizon $T$.

\subsection{Maximum Ramp Rates}
If the production of renewable energy sources changes rapidly due to weather events, these power ramps must be compensated.
Therefore, the ability of traditional generation plants and storage facilities to follow these ramps plays a decisive role in system reliability and cost efficiency.
In order to avoid fluctuations on the generation side, the physical limitations of the corresponding plants are taken into account in the optimization.
In addition power ramp constraints can be applied to network connection points to limit the up and down ramps at this point and to compensate and smooth the power variations. The following equations represent the restrictions for up and down ramps ($R^{U}_{g}, R^{D}_{g}$) of the respective unit $g$ \cite{morales-espana_tight_2013, morales-espana_mip_2014}:
\begin{align}
 p_{g,t}-p_{g,t-1} &\leq R^{U}_{g} \hspace{0.6cm}\forall{g},\forall{t} \\[1em]
 p_{g,t-1}-p_{g,t} &\leq R^D_{g}\hspace{0.6cm}\forall{g},\forall{t}
\end{align}

\subsection{Minimum Up and Down times }
Each unit $g$ has limitations regarding its minimum runtime (uptime, $T^{U}_{min,g}$) and downtime ($T^{D}_{min,g}$). 
For example, a CHP should be operated at the optimal operation point as long as possible and not be switched on or off arbitrarily.
According to Rajan and Takriti \cite{Rajan2005MinimumUP} these restrictions can be formulated as follows:
\begin{align}
u_{g,t} - u_{g,t-1} &= v_{g,t}-w_{g,t} \\[1em]
\sum_{\mathclap{i=t-T^{U}_{min,g}+1}}^{t}\; v_{g,i}\quad &\leq u_{g,t} \\[1em]
\sum_{\mathclap{i=t-T^{D}_{min,g}+1}}^{t}\; w_{g,i}\quad &\leq 1-u_{g,t}
\end{align}

Here $v_{g,t}$ and $w_{g,t}$ are binary variables that indicate the start up and shutdown process of the respective system.
The binary variable $u_{g,t}$ describes the operating state of the corresponding installation.

\section{Uncertainty Modelling}
\label{sec:uncertainty_model}

\begin{figure}[tb]
  \centering
  \includegraphics[width=.45\textwidth]{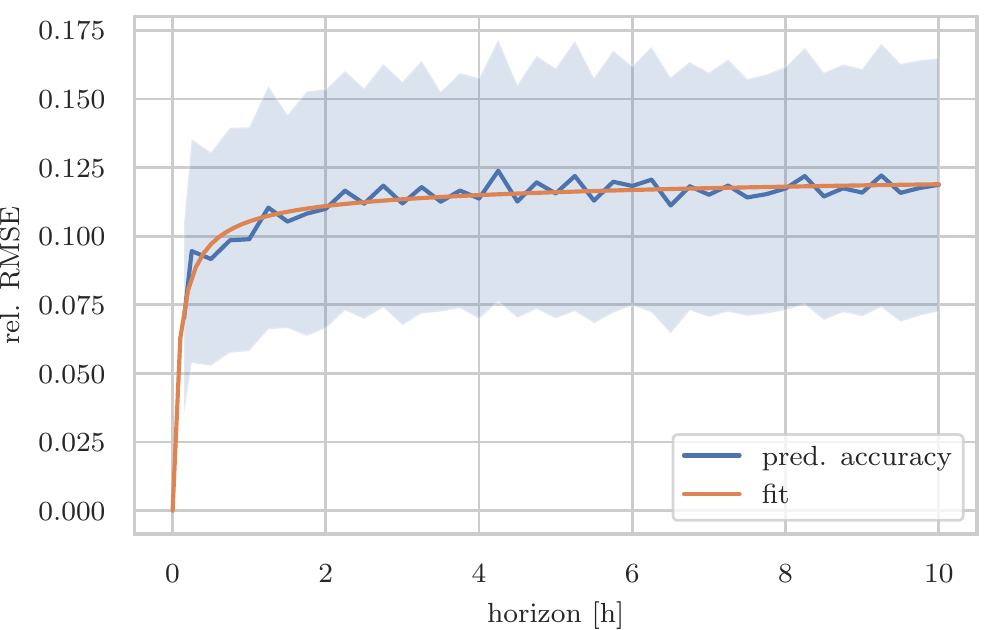}
  \caption{The relative prediction RMSE over the horizon and the model fit.}
  \label{fig:pred_err_fit}
\end{figure}

The problem of investment decisions in the presence of uncertainty was one of the first applications where uncertainty was taken into account within optimization processes. 
This led researchers to capture uncertainty in modeling energy resource allocation. Thus, there is a transmission from traditional deterministic approaches to stochastic and robust optimization \cite{wytock_dynamic_2017} with the aim to answer the demands on system reliability.
In this paper a robust optimization approach facilitating a worst case prediction error assumption is implemented.

\subsection{Load and Generation Predictions}

In the recent years load and generation predictions have been predominantly generated using machine learning techniques \cite{felder_wind_2010, yaici_artificial_2015}.
The predictions for this paper are generated with artificial neural networks using various weather models as input data. 

In order to generate a worst case optimization scenario for the optimization process, the accuracy of the load and generation prediction has to be estimated. 
Therefore a model for the prediction accuracy is created by fitting an inverse harris function to the known prediction errors. The prediction errors are calculated based on data coming from a large monitored PV-farm. 

In Fig.\,\ref{fig:pred_err_fit} the relative prediction error depending on the corresponding forecast horizon is shown.
These samples can be approximated with the following function:

\begin{equation}
    \varepsilon = t / (a + b t^c)
    \label{eq:fit_fun}
\end{equation}
where $\varepsilon$ denotes the predictin error and $t$ the forecast horizon. The parameters $a$, $b$ and $c$ are fitted to the data in a least squares sense.

 \begin{figure}[b]
  \centering
  \includegraphics[width=.3\textwidth]{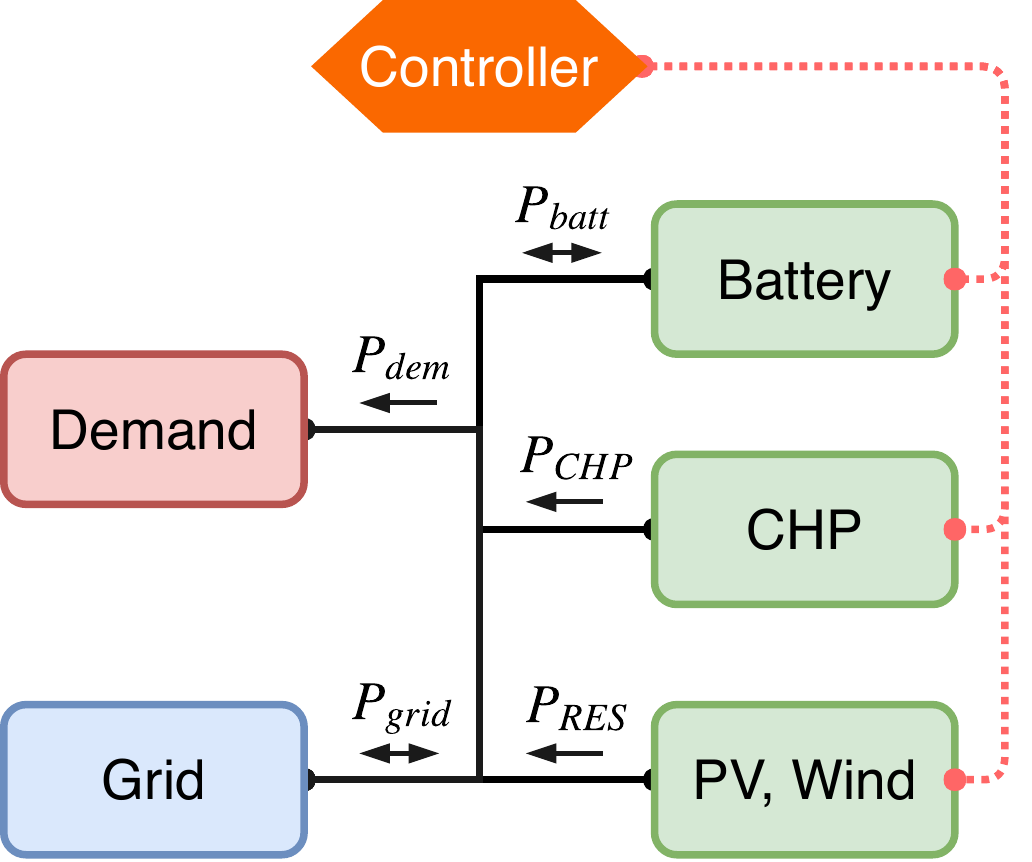}
  \caption{Sketch of the simulation scenario with the corresponding units $g$ and the power flows.}
  \label{fig:vpp_graphic}
\end{figure}

\section{Simulation Scenario}

The topology of the virtual power plant used in this simulation study is shown in Fig. \ref{fig:vpp_graphic}.
The corresponding network model consists of power nodes representing the distribution-energy-resources (DER), the local demand and the external grid. The DERs comprise the CHP and the renewable energy sources such as PV and Wind.
Furthermore, a battery serves as a short term energy storage to balance generation and demand. It is assumed that the battery charge power ($P_\text{batt}$) and the electrical generation by the CHP ($P_\text{CHP}$) can be controlled by the energy management algorithm.
The CHP is assumed to have a minimum operating power during operation. 
Moreover it is assumed that the renewable energy generation can be curtailed ($P_\text{RES}$).
To capture the seasonal variability of load and generation without increasing the simulation complexity, three sample days are simulated: a summer day, a transitional day between summer and autumn and a winter day. The complete list of constraints for this scenario is shown in Table \ref{tab:scenario}.

 \begin{table}[htb]
	\caption{Simulation Scenario and Constraints.}
	\centering
	\begin{tabular}{l r}
		\hline
		CHP startup time & 15\,min. \\
		CHP min. uptime & 20\,min. \\
		CHP min. downtime & 15\,min. \\
		$\overline{P}_\text{CHP}$     & \SI{20}{\kilo\watt}  \\
		$\underline{P}_\text{CHP,\, on}$    & \SI{06}{\kilo\watt}  \\
		$\overline{P}_\text{RES}$     & \SI{30}{\kilo\watt}  \\
		$\overline{P}_\text{demand}$  & \SI{50}{\kilo\watt}  \\
		$\overline{P}_\text{grid}$    & \SI{10}{\kilo\watt}  \\
	    $C_\text{batt}$             &    \SI{20}{\kilo\watt\hour}  \\
		$\overline{P}_\text{batt}$    & \SI{20}{\kilo\watt}  \\
		$t_\text{control}$ 		& $[0, 5, 10, 15, 30, 60, 90, 120, \dots , 600]$\,min. \\
		$\Delta t_\text{opt}$ & 5\,min.\\
		\hline
	\end{tabular}
	\label{tab:scenario}
\end{table}

\subsubsection{Startup Times}
Each system is subject to start-up and shut-down restrictions. If power generation plants with relatively long start-up times are available in the virtual power plant, it is therefore more important to obtain a good forecast so that the corresponding processes and machines can be started up in time.

\begin{figure}[tb]
	\centering
	\includegraphics[width=.45\textwidth]{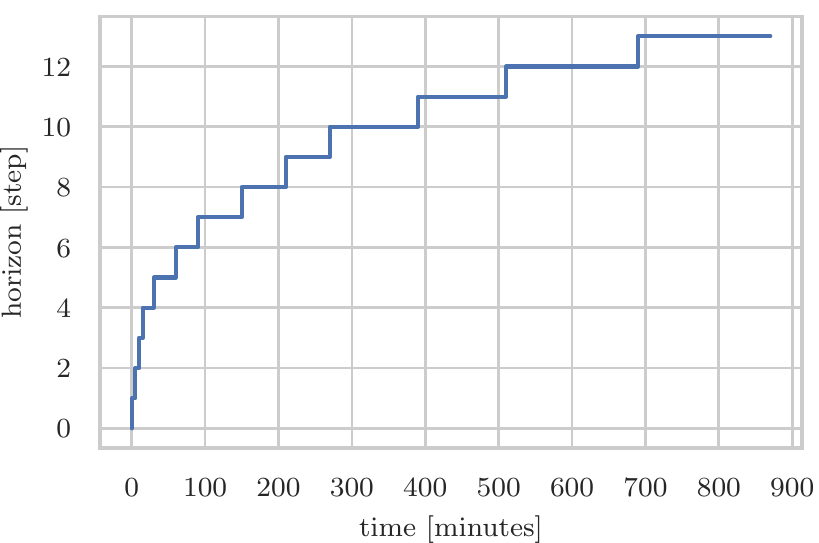}
	\caption{The $\Delta t$ of each timestep increases for larger horizons. This enables to use the accuracy of the short term predictions and to capture the energy balance of the longer horizons in the optimization.}
	\label{fig:horizon}
\end{figure}

\subsubsection{Horizon and Timesteps}
In order to make optimal use of the information from the short term forecast and at the same time take the energy balance of the coming hours into account, a variable increment of $\Delta t$ over a horizon of 15h is used. For the first control points $t_{\text{control}}$ a step size of 5 minutes is chosen, which is then increased to 3 hours (see Fig.\,\ref{fig:horizon}).
This also reduces the dimension of the optimization problem, which speeds up the computation time.
The optimization is run every 5 minutes ($\Delta t_\text{opt}$).

\subsection{Cost Assumptions}

For the optimization process it is necessary to define the costs for each DER as well as the electricity and the battery costs. These costs are summarized in Table \ref{tab:costs}.
For the battery it is assumed that only the capital and the operational costs contribute to the optimization costs (taxes are neglected)\cite{lazard_2017}. For the electrical generation from the CHP the costs are assumed to be \eurkwh{.10} \cite{nguyen_levelized_2014}. 

The electricity price is assumed to be  \eurkwh{.30} and the feed-in tariff  \eurkwh{.03}. In addition costs for the limitation of renewable generation are applied corresponding to the levelized cost of electricity \cite{kost_levelized_2018}. 

\begin{table}[htb]
	\caption{Costs used for the optimization model.}
	\centering
	\begin{tabular}{l l}
		\hline
		$\kappa_\text{batt}$              & \eurkwh{.15}            \\
		$\kappa_\text{CHP,\,gen}$         & \eurkwh{.10}            \\
		$\kappa_\text{CHP,\,start}$       & \SI{.30}{\eur} p. start \\
		$\kappa_\text{grid,\,draw} $      & \eurkwh{.30}            \\
		$\kappa_\text{grid,\,feedin}$     & \eurkwh{.03}            \\
		$\kappa_\text{RES,\,curtailment}$ & \eurkwh{.10}            \\ \hline
	\end{tabular}
	\label{tab:costs}
\end{table}

\begin{figure}[b]
	\centering
	\includegraphics[width=0.9\linewidth]{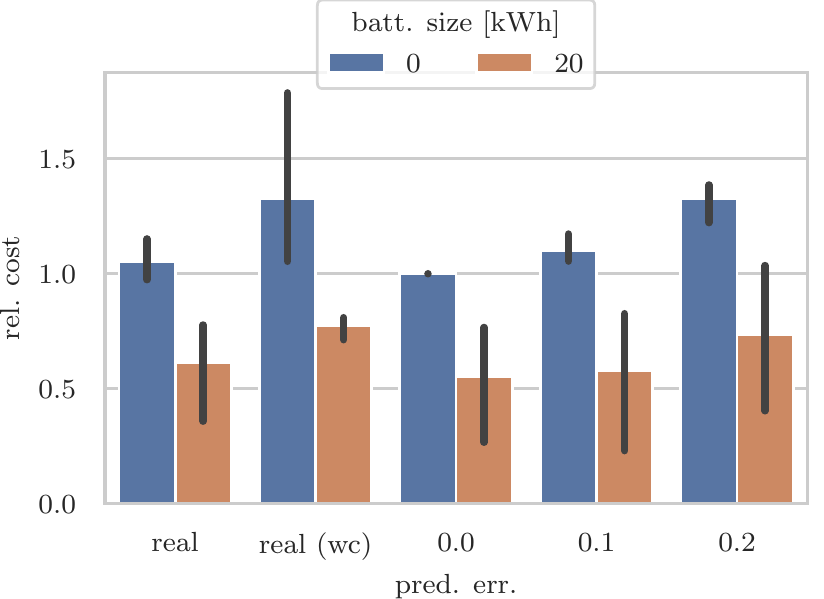}
	\caption{Comparison of the absolute costs of the different error scenarios with and without battery. The error bars show the \SI{95}{\percent} confidence intervall of the underlying data.}
	\label{fig:batt_comp}
\end{figure}

\begin{figure}[tb]
	\centering
	\includegraphics[width=.9\linewidth]{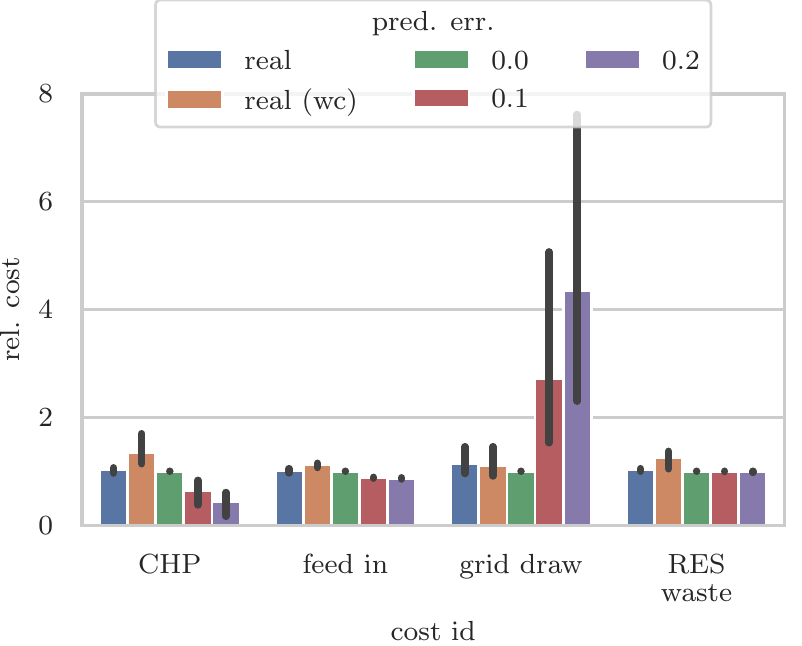}
	\caption{Comparison of the relative costs per device for the case without battery.}
	\label{fig:cost_comp_nobatt}
\end{figure}

\section{Results}

To show the influence of the prediction error on scheduling applications a simulation based on the before mentioned scenarios is carried out. All results are normalized to the base case: perfect predictions ($\varepsilon_\text{pred}=0.0$) and no battery.

In Fig.\,\ref{fig:batt_comp} the overall costs depending on the prediction accuracy for a scenario with and without battery are shown. On the x-axis the prediction error scenarios are listed and on the y-axis the costs relative to a scenario without prediction error nor battery are displayed. The first error scenario shows the costs based on real predictions generated with neural networks. The second item describes a worst case scenario based on real predictions (robust optimization). This is the most conservative scenario which is supposed to ensure a save operation of the plant in presence of prediction errors. The following columns describe synthetic prediction scenarios with a RMSE of $0.0$, $0.1$ and $0.2$ relative to the maximum value of the profile.
The error was modeled using the fit described in equation (\ref{eq:fit_fun}) normalized to the maximum power of the corresponding unit.

\subsection{Sensitivity on the prediction error}

\begin{figure}[b]
	\centering
	\includegraphics[width=.9\linewidth]{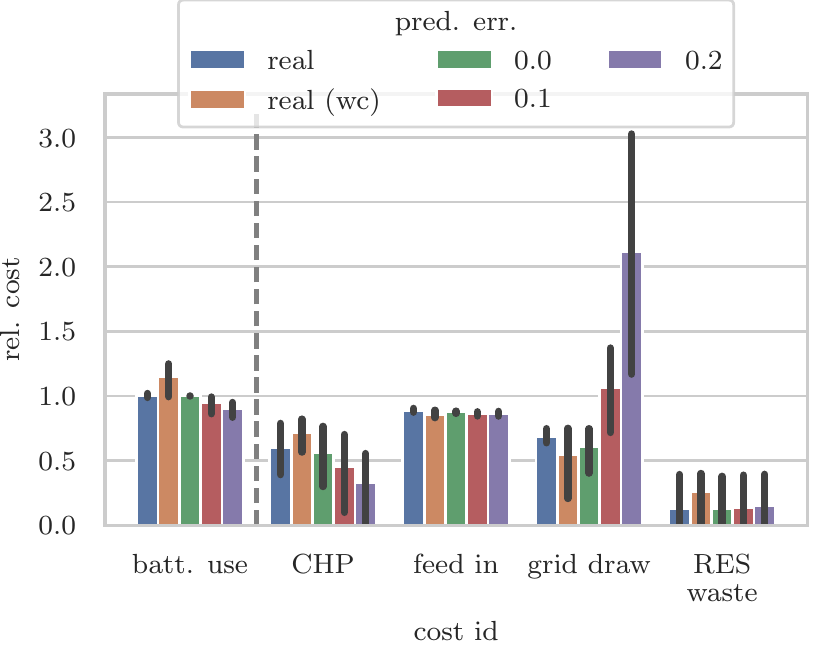}
	\caption{Comparison of the relative costs per device for the case with a battery. All costs are rekative to the reference case except the battery costs: they are normalized to the scenario with perfect prediction and a battery. A enlarged version is provided in Fig\,\ref{fig:cost_comp_zoom}.}
	\label{fig:cost_comp}
\end{figure}

\begin{figure}[t]
	\centering
	\includegraphics[width=.9\linewidth]{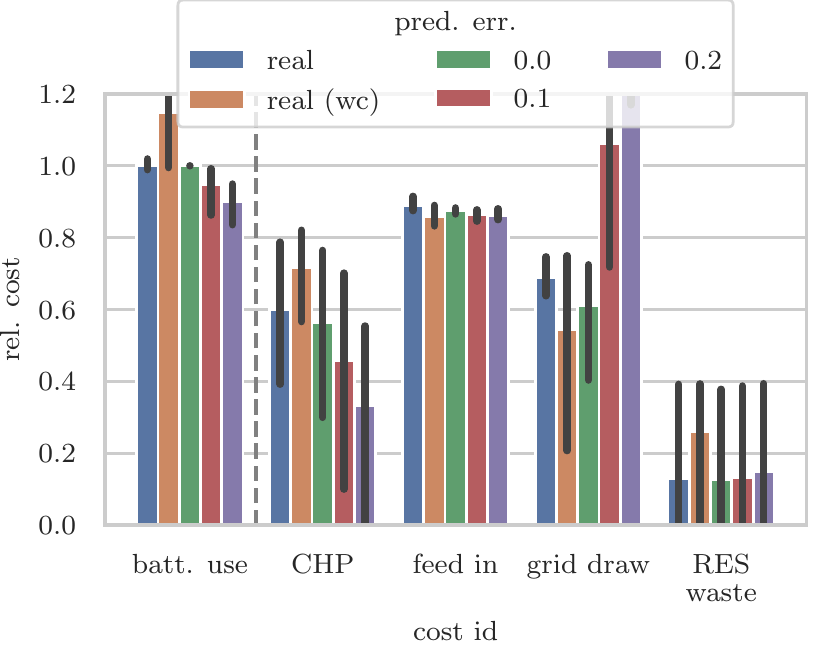}
	\caption{Enlarged version of Fig\;\ref{fig:cost_comp}. Comparison of the relative costs per device for the case with battery.}
	\label{fig:cost_comp_zoom}
\end{figure}

In Fig.\,\ref{fig:batt_comp} it can be observed that the costs increase with a decreasing prediction accuracy. 
One reason for this is due to the delayed start-up behavior of the CHP unit: The controller must request the required power 15 minutes before the call time so that there is enough time for the co-generation unit to start up.
This can also be seen in Fig.\,\ref{fig:cost_comp_nobatt}, where the costs per device for a scenario without a battery are shown. 
The use of the CHP declines with increasing prediction error. At the same time the amount of energy drawn from the grid increases to compensate the shortage of energy from the CHP.
The amount of curtailed renewable generation (RES waste) is about the same for all scenarios.

The same plot is shown in Fig\,\ref{fig:cost_comp} for a scenario with a 20kWh battery. Compared to the scenario without a battery very little RES energy has to be curtailed as the excess can be stored in the battery. 

The battery usage itself decreases with an increasing prediction error.
As an increased battery usage also raises the operation costs it is only used when it is predicted to be beneficial for the total costs over the optimization horizon. If these predictions overestimate the local generation and underestimate the local demand, the battery usage decreases.

In all cases the use of a battery decreases the overall costs as the battery round trip price per kWh is in this scenario cheaper than buying electricity from the grid. 
Furthermore, the battery can compensate the effect of prediction errors to some extend: the costs for $\varepsilon_\text{pred}=0.1$ are almost the same as for the case with $\varepsilon_\text{pred}=0.0$ and the required power supply from the mains can be reduced.

The real predictions perform around \SI{5}{\percent} worse compared to the perfect predictions ($\varepsilon_\text{pred}=0.0$). The worst case scenario creates the highest costs but provides also the highest operational safety.

\subsection{Robust optimization}
	
For the robust optimization the worst case estimate for the prediction error is used for the optimization. This ensures a failsave operation despite the presence of prediction errors.
In Fig.\,\ref{fig:cost_comp} the costs per device relative to the case with perfect predictions are shown.
The worst case prediction shows the largest CHP usage and the least grid draw. This is due to the worst case optimization: as the CHP exhibits a deadtime of 15 minutes it will be switched on if the predicted profiles show any sign of a negative energy balance which could exceed the grid limits.
This leads also to the larger use of the battery (see Fig.\,\ref{fig:cost_comp_zoom}), as the all excess of energy is stored in the battery before fed into the grid. 
Furthermore, the amount of energy supplied by the grid is less than for any other strategy, which shows the robustness of the approach. One further drawback is the increased amount of curtailed renewable energy.

\section{Conclusion}

The results presented in this paper show the importance of accurate forecasts for smart-grid or micro-grid operations. 
For the presented scenario without a battery a relative prediction error of \SI{10}{\percent} leads to an \SI{8}{\percent} increase in operational costs and a \SI{20}{\percent} prediction error leads to an increase of about \SI{25}{\percent}. If a battery is added to the system a relative prediction error of \SI{10}{\percent} leads approximately to an \SI{2}{\percent} increase in operational costs and a \SI{20}{\percent} error leads to an increase of about \SI{30}{\percent} respectively.

Especially if devices with a start-up time delay are included in the system, an accurate short term power prediction is needed.
The effects of the prediction error can be reduced if a battery is used.
If a large error margin is required (e.g. for critical facilities running in an environment with tight grid constraints) a robust optimization can be performed. This increases the failure safety of the operation but also increases the costs.

Also for storage operation a high quality forecast is important to use the battery efficiently.